\documentclass[conference]{IEEEtran}
\IEEEoverridecommandlockouts
\usepackage{cite}
\usepackage{amsmath,amssymb,amsfonts}
\usepackage{algorithmic}
\usepackage{graphicx}
\usepackage{tikz}
\usepackage{textcomp}
\usepackage{xcolor}
\def\BibTeX{{\rm B\kern-.05em{\sc i\kern-.025em b}\kern-.08em
    T\kern-.1667em\lower.7ex\hbox{E}\kern-.125emX}}

\newcommand\copyrighttext{%
  \footnotesize © 2024 IEEE.  Personal use of this material is permitted.  Permission from IEEE must be obtained for all other uses, in any current or future media, including reprinting/republishing this material for advertising or promotional purposes, creating new collective works, for resale or redistribution to servers or lists, or reuse of any copyrighted component of this work in other works.}
\newcommand\copyrightnotice{%
\begin{tikzpicture}[remember picture,overlay]
\node[anchor=south,yshift=10pt] at (current page.south) {\fbox{\parbox{\dimexpr\textwidth-\fboxsep-\fboxrule\relax}{\copyrighttext}}};
\end{tikzpicture}%
}

\begin{document}

\title{Active Islanding Detection Using Pulse Compression Probing\\}

\author{\IEEEauthorblockN{Nicholas Piaquadio, N. Eva Wu, Morteza Sarailoo}
\IEEEauthorblockA{\textit{Department of Electrical and Computer Engineering} \\
\textit{Binghamton University}\\
Binghamton, NY USA\\
\{npiaqua1,evawu,msarail15\}@binghamton.edu}
}

\maketitle

\begin{abstract}
An islanding detection scheme is developed using pulse compression probing (PCP). A state space system realization is taken from the probing output. The nu-gap metric is applied to compare the measured system to fully intact system and classify it as islanded, or grid-connected. The designed detector displays fast operation, accurate islanding detection results under varying grid condition, and is physically implementable at the terminals of an inverter. The method is verified via electro-magnetic transient (EMT) simulation on a modified IEEE 34 bus test system with randomized loads and simultaneous probing at three independent solar plants, with the probing signal directly implemented into the logic of a switching inverter model.
\end{abstract}

\copyrightnotice

\vspace{-3mm}
\section{Introduction}

The rapid growth of distributed energy resources (DER), small generation facilities typically connected at distribution level and through an inverter, has created a need for more advanced solutions to the islanding detection problem. In this work, an active islanding detection scheme is implemented via power system probing \cite{Chakraborty_2022}, injecting a signal into the grid to extract useful information.

The goal of an islanding detector is to determine whether a facility is connected to the broader transmission network (grid-connected) or has been isolated by a switching event or fault (islanded) \cite{Bower_2002}. Prompt detection and disconnection from an islanded state is necessary for utility worker safety, and to prevent equipment damage. However, false trips, or disconnection from a grid-connected state, compromise resource availability and impact grid reliability. To this end, many islanding detection methods have been studied and implemented in recent years \cite{Bekhradian_2022}. 

Passive islanding detection schemes, or passive schemes, measure grid quantities such as voltage, current, and frequency to determined if a DER is grid connected, or islanded \cite{Chowdhury_2008}. Passive schemes vary greatly in design and implementation. Relay-based schemes detect conditions like under voltage, over-voltage, and off-nominal frequency. More recent developments include the use of machine learning \cite{Baghaee_2020}, and wavelet transforms to decompose time-domain grid data during a disturbance \cite{Prakash_2012}.

Regardless of their level of sophistication, all passive schemes have a 'non-detection zone', or a range of islanded conditions they cannot detect, especially the case of a 'balanced island' where the islanded generation is roughly equal to the islanded load \cite{Bower_2002}. False tripping due to disturbances such as load and capacitor switching and DER startup harmonics is also problematic \cite{Prakash_2012}. 

Active islanding detection schemes inject a signal into the grid improve detector accuracy and reduce the non-detection zone. One of the most studied active schemes is the Sandia Frequency Shift (SFS) \cite{Ropp_1999}. SFS is designed to cause a frequency perturbation in islanded conditions, and does so by introducing a small positive feedback in active power. 

In this work, islanding detection is achieved via power systems probing, which has found applications ranging from swing-mode identification to grid topology detection \cite{Chakraborty_2022}. The majority of these applications inject a signal as a phasor, averaged over several cycles, such as real or reactive power. The authors' previous work involved the use of Pulse Compression Probing (PCP), \cite{Wu_2014}, which measures the impulse response between the system input, where the probing signal is injected, and system output, where the response is measured.

In \cite{Piaquadio_2023}, the probing signal was injected as high-frequency voltage signal, and the resulting current was measured. The impulse response was then processed to yield a small-signal model of the IEEE 13 node distribution test feeder. The advantage of probing in this manner lies in the high speed: the entire probing process in \cite{Piaquadio_2023} was completed in under 7 cycles in a 60 Hz power system, or 117 ms. 

This paper builds on the concept of probing via PCP to develop an active islanding detection scheme. The goal is to leverage the properties of PCP that are desirable for a detector. Foremost, the time-domain probing process allows for rapid detection. Secondly, having the impulse response of the system as seen by the DER allows the application of robust control concepts, which are used to minimize the non-detection zone, and guarantee that the detector can operate in changing system conditions. Finally, the design of the probing signal allows injection directly from a DER facility, and for multiple DER to probe the system simultaneously. This process is furthermore local to each plant, and avoids the cooperation and point to point communications infrastructure needed for methodologies like power line carrier and direct transfer trip. These properties are demonstrated via simulation on a modified version of the IEEE 34 bus test system \cite{Kersting_2001}.

The remainder of the work is outlined as follows. Section II provides a background on the working principles of PCP, and the design of the islanding detector. Section III describes the test system used, and Section IV demonstrates the effectiveness of the method via simulation. Conclusions and future work directions are contained in Section V.

\section{Probing Method and Gap Metric}

This section provides a review of the working principles of Pulse Compression Probing (PCP) \cite{wu_2011}. From the probing output, an islanding detection scheme is developed by applying the nu-gap metric, a measure of the distance between two systems \cite{Vinnicombe_1993}. 

\subsection{Pulse Compression Probing}

PCP introduces a probing signal to a given system, and measures that system's small-signal impulse response, linearized about its operating point. The construction of a probing signal for this purpose follows the same steps as \cite{Piaquadio_2023}. A brief overview is given here. 

\begin{figure}[h]
    \centering
    \includegraphics[scale=0.32]{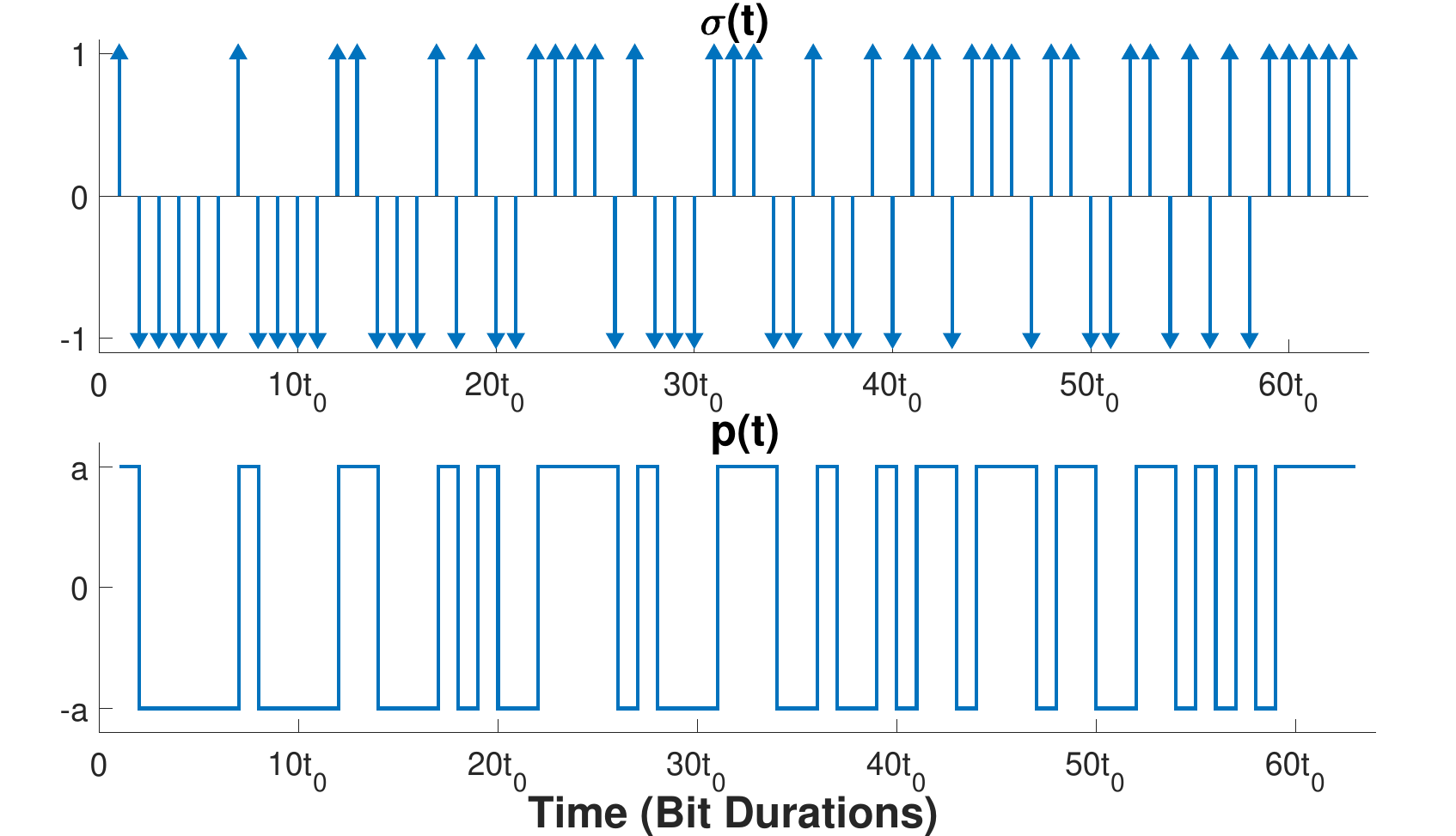}
    \caption{One period of $\sigma(t)$ \cite{Piaquadio_2023}, an impulse-based pseudo-random binary sequence (PRBS) \cite{Ljung_1999} of order n = 6, and its corresponding continuous time signal, $p(t)$, a pseudo-random binary pulse train (PRBPT). The PRBPT is used as the probing signal in this work. }
    \label{Piaquadio_1}
\end{figure}

The probing signal is based on a Pseudo-Random Binary Sequence (PRBS) \cite{Ljung_1999}. A PRBS, $\sigma (t)$, composed of positive and negative impulses, is shown in the top plot of Fig. \ref{Piaquadio_1}. For a given order, $n$, the PRBS has $2^n -1$ impulses.

To construct the signal shown in the lower half of Fig. \ref{Piaquadio_1}, the PRBS is convolved with a rectangular pulse, $\eta (t)$, with unit magnitude and width $t_0$. The result is multiplied by a scaling factor $a$, as shown in (\ref{probe_sig}). 

\begin{equation}
    p(t) = a \eta(t) * \sigma(t)
    \label{probe_sig}
\end{equation}

Here, $p(t)$ is called a Pseudo-Random Binary Pulse Train (PRBPT) \cite{wu_2011}. The width of $\eta (t)$, $t_0$, gives the "bit duration" of the PRBPT. The total period of the probing signal is $t_0$ times the number of pulses, or $T_p = (2^n-1)t_0$. $T_p$ should be selected at least as long as memory length of probed system. In other words, PCP captures the impulse response out to $T_p$ seconds. In a typical design, $t_0$ is selected based on the needed sampling rate and the order, $n$,  is adjusted to set $T_p$.

\begin{figure*}[t]
    \centering
    \includegraphics[scale=0.62]{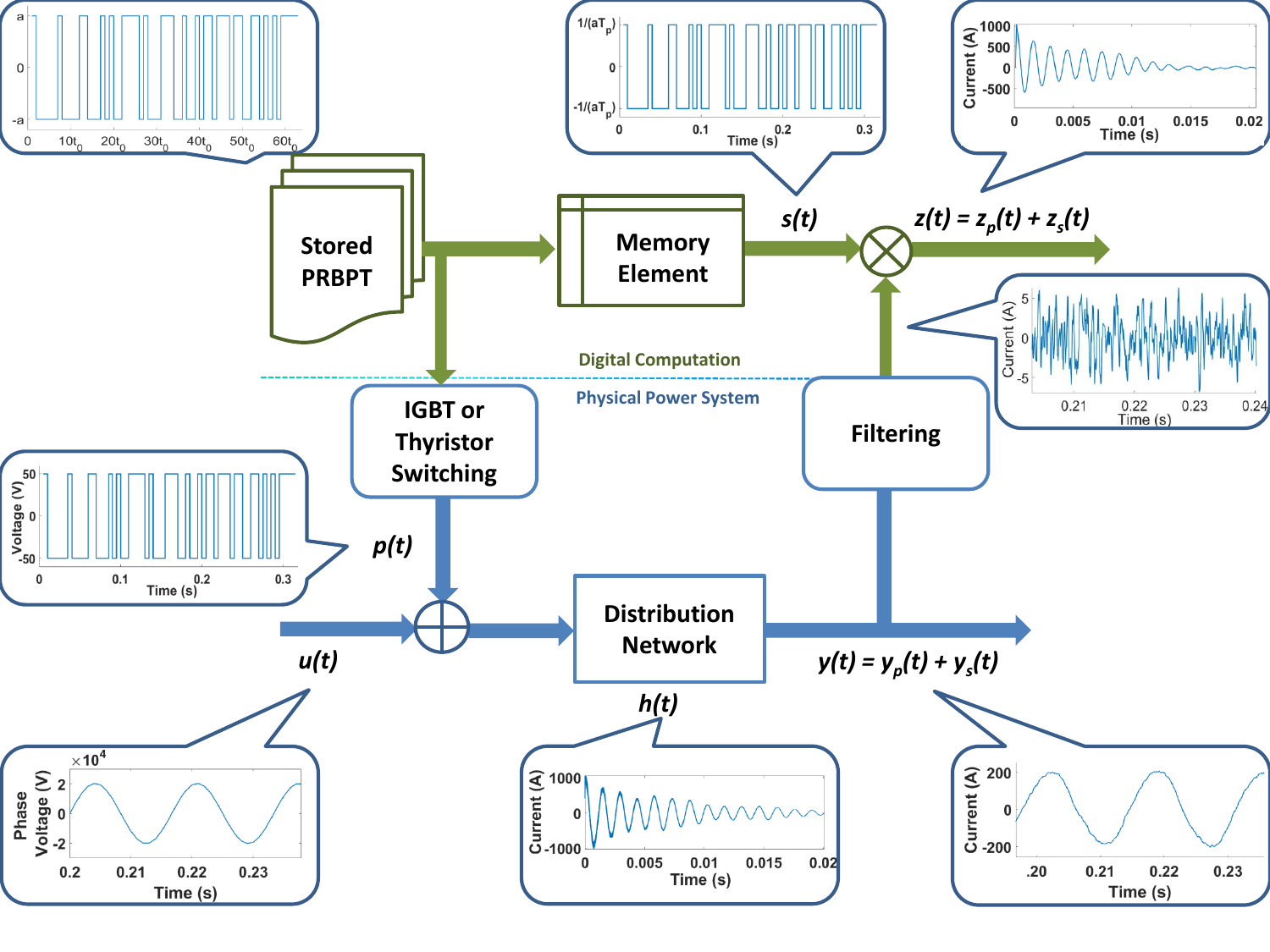}
    \vspace{-0.4cm}
    \caption{A schematic showing the process of pulse compression probing (PCP) and example signals. A power electronics device injects the probing signal $p(t)$ as a voltage added, $u(t)$, the bus voltage created by other sources in the network. The measured system response (current) is cross correlated with $s(t)$, a repetition of the original probing signal to yield the output.}
    \label{Piaquadio_2}
     \vspace{-0.3cm}
\end{figure*}

The PRBPT is used to monitor the distribution network as shown in Fig. \ref{Piaquadio_2}. $p(t)$ is implemented in inverter-switching logic as a voltage signal. It is added to $u(t)$, the bus voltage due to other sources. $y(t)$, is the measured inverter current and is broken into two components: $y_p(t)$, the current caused by the probing signal, and $y_s(t)$, the current caused by $u(t)$.

The system output, $y(t)$ is filtered and cross correlated with a reference signal $s(t)$. $s(t)$ is a cyclic (looped) repetition of probing signal $p(t)$ with magnitude $1/a T_p$. The result of this cross correlation is the probing output, $z(t)$. $z(t)$ is expressed as two components: $z_p(t)$, the output due the probing signal, and $z_s(t)$, the output due to $u(t)$. It is shown in \cite{wu_2011} that as the bit duration $t_0$ becomes sufficiently small, $z_p(t)$ approaches the system impulse response, $h(t)$  (\ref{xcorr2}):

\begin{equation}
    z_{p}(t) = h(t)*p(t)\otimes s(t) \approx \sum_k  h(t-kT_p) 
    \label{xcorr2}
\end{equation}

Where $\otimes$ represents cross correlation. The essence of PCP signal design is to select parameters $a$, $n$, and $t_0$, such that: 

1. The probing output is dominated by the impulse response $z(t) \approx z_{p}(t)$.

2. The system output is dominated by the nominal output $y(t) \approx y_{s}(t)$. 

\subsection{Detector Design}

The goal of this work is implement PCP in an electrical distribution network and  perform islanding detection. The output of PCP is the small-signal impulse response of the network from the probing terminal. Therefore, concepts from control systems theory are applied to develop a detector.

To perform islanding detection, the probed impulse response is compared against a nominal impulse response, assuming the system is intact, or has all lines in service. Rather than compare impulse responses directly, system realization, the process of extracting a control systems model from measured data \cite{Zhou_1999}, is used to obtain the state-space model of each system.

The probing output $z_p(t)$ is an estimate of the discrete-time impulse response of the network, and therefore an estimate of its Markov Parameters, the sequence of sampled output values obtained when a state-space model is excited by an impulse \cite{Chen_1999}. The Hankel matrix is estimated following \cite{Piaquadio_2023}, and a state-space realization $\{A,B,C\}$ is then taken using the Eigensystem Realization Algorithm (ERA) \cite{Jaung_1985}. 

 The nu-gap metric, or $\delta_{\nu}$ \cite{Vinnicombe_1993}, is selected to compare the resulting state space systems. The nu-gap metric ranges from 0 to 1 and compares the controllability of two closed-loop networks. If two state space systems are close in the nu-gap metric, there is a large set of controllers that can stability both. If the systems are far apart, there are few or no such simultaneous stabilizing controllers \cite{Zhou_1999}.  $\delta_{\nu}$ can be computed point by point in the frequency domain for two systems, $P_1(j\omega$), and $P_2(j\omega$), according to (\ref{nugap2}) \cite{Vinnicombe_1993}. 

 \begin{equation}
 \begin{split}
    &\psi(P_1(j\omega),P_2(j\omega)) = \frac{|P_1(j\omega)-P_2(j\omega)|}{\sqrt{1+|P_1(j\omega)|^2}\sqrt{1+|P_2(j\omega)|^2}} \\
    &\delta_{\nu} = \max_{\omega} \psi(P_1(j\omega),P_2(j\omega))
\end{split}
 \label{nugap2}
 \end{equation}


In this work, we probe a power system via a voltage signal and measure the injected current. $P(j\omega)$ is therefore a measurement of the Thevenin impedance (or admittance) seen from the probing point at a particular frequency: $P(j\omega) = Z(j \omega) \approx R +j\omega L$. At zero frequency, this is a measurement of resistance only, and at high frequency, this is roughly a measurement of reactance only. As $\delta_\nu$ is the maximum over all frequencies, the method is adaptable to networks of differing R/X ratio. In other words, the best weighting between R and X, that maximally separates the compared systems, is automatically selected by the metric.

The nu-gap was chosen based on this principle and the following practical consideration. Many voltage controllers are stable for all grid-connected conditions, as the connection to the transmission grid acts as a strong source\cite{Saadat}. This strong source regulates voltage throughout the distribution feeder, including the inverter terminals. Therefore, when comparing two grid-connected systems, there will be many simultaneous stabilizing controllers, and it is hypothesized that grid-connected conditions will be close in the nu-gap.

In an islanded condition, the connection to the transmission grid is broken by a line outage. There are expected to be large number of controllers that would stabilize the inverter when grid-connected, but would be unstable when islanded.  Hence, the family of simultaneous stabilizing controllers when comparing an islanded condition to a grid-connected condition is expected to be small, and the nu-gap metric should be large.

This is, in essence, a similar concept to Sandia Frequency Shift \cite{Ropp_1999}, which designs a frequency controller that will be stable when grid-connected, and become unstable when islanded. Rather than implement a single controller, which could reach instability inadvertently, using the nu-gap mathematically evaluates the family of all possible controllers.

\section{Test System and Simulation Setup}

To demonstrate the usage of PCP in a distribution network, the IEEE 34 bus test system \cite{Kersting_2001} is modified to add three three-phase solar farms, each represented by an aggregated switching inverter model, filter, and step-up transformer \cite{Giroux_2012}. The 34 bus test system represents a 24.9 kV feeder in Arizona, United States. It is unbalanced, and includes a small 4.16 kV network.

\begin{figure}[t]
    \centering
    \includegraphics[scale=0.33]{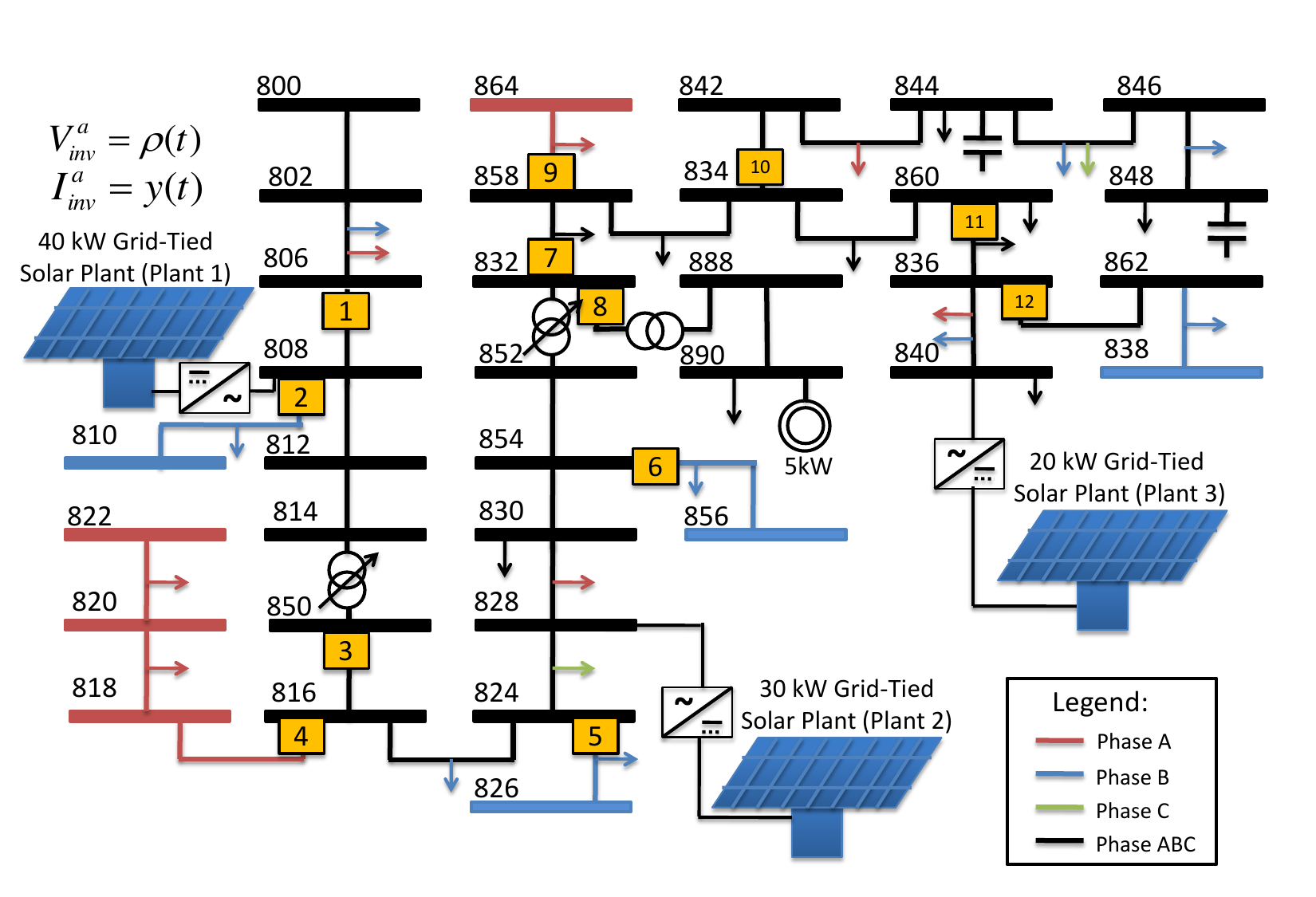}
    \vspace{-4mm}
    \caption{A modified version of the IEEE 34 bus test system \cite{Kersting_2001}. Three distribution-scale, grid-tied solar plants, breakers, and a 5 kW induction motor load are added. Phases are color coded, and breakers are indicated by numbered yellow boxes. PCP is assumed to be available at each solar plant. Probing signals are labeled at Plant 1.}
    \label{Piaquadio_3}
     \vspace{-4mm}
\end{figure}

A 5 kilowatt induction motor load is additionally added at Bus 890, and breakers/reclosers are assumed at the locations indicated in Fig. \ref{Piaquadio_3}. 

\subsection{Implementation of the Probing Signal}

To inject PRBPT signals from the inverters, an override sequence was built into the logic that generates firing pulses for the 3-phase H bridge. The PRBS is injected by transitioning between positive and negative DC voltage at the switching frequency of the inverter, which is assumed to be 25 kHz. This frequency was needed for adequate resolution on the network response, and is beyond the 20 kHz switching frequency of today's standard medium-voltage (600V) Insulated Gate Bipolar Transistors (IGBTs), but is likely achievable with near-future IGBTs \cite{Zhang_2019} or metal-oxide-semiconductor field-effect transistor (MOSFET) based inverters. 

 Table \ref{PCP_Table} shows the probing signal parameters used in simulation. Each plant has a unique bit duration such that the probing signals of the other two are averaged out in the cross correlation step of probing (see Fig. 2), eliminating interference between DER plants.

 \begin{table}[h!] 
    \centering
    \vspace{-3mm}
            \caption{Parameters of Selected Probing Signal}
    \begin{tabular}{||c|c|c||}
    \hline
        Probing Parameter & Symbol &Value  \\
           \hline
            Bit Duration (Plant 1) & $t_0$ & 40 $\mu s$\\
             \hline
            Bit Duration (Plant 2) & $t_0$ & 48 $\mu s$\\
             \hline
            Bit Duration (Plant 3) & $t_0$ & 56 $\mu s$\\
             \hline
             PRBPT Order & $n$ & 12\\
             \hline
             PRBPT Magnitude & $a$ & 600 Volts\\
             \hline
    \end{tabular}
    \label{PCP_Table}
    \vspace{-3mm}
\end{table}

\begin{figure}[t]
    \centering
    \includegraphics[scale=0.41]{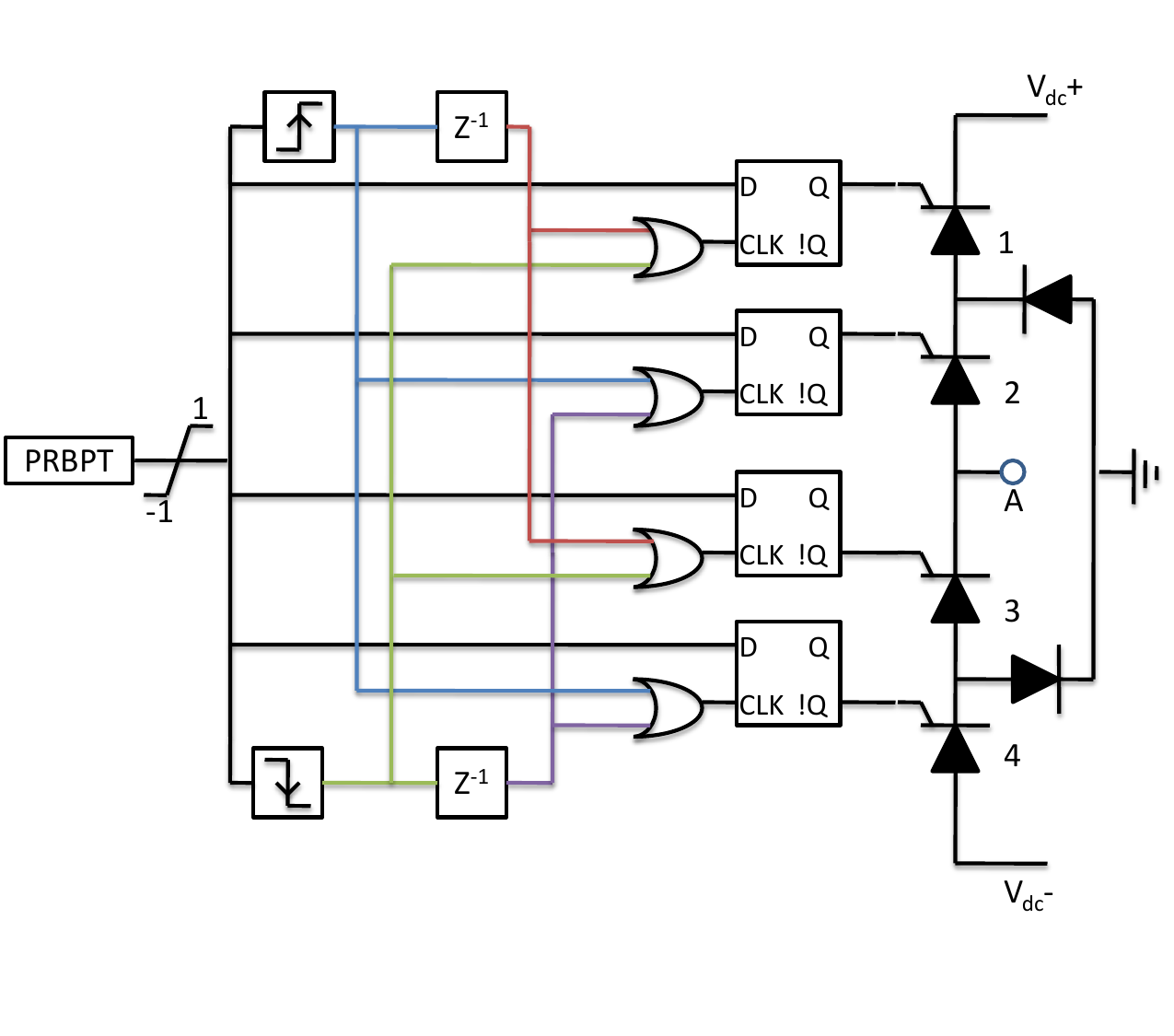}
    \vspace{-14mm}
    \caption{Implementation of a PRBPT in the firing sequence of a 4-switch H-bridge with neutral. On a rising edge, the first signal (blue) closes switch 2, and opens switch 4, and a delayed signal (red), closes switch 1 and opens switch 3. On a falling edge, the first signal (green) opens switch 1 and closes switch 3, while a delayed signal (purple) closes switch 4 and opens switch 2.}
    \label{Piaquadio_4}
    \vspace{-4mm}
\end{figure}

This results in probing periods, $T_p$, of 164 ms, 197 ms, and 223 ms, respectively, for solar plants 1, 2, and 3. The PRBPT signal is implemented into inverter switching logic via the scheme shown in Fig. \ref{Piaquadio_4}. 

 From the figure, the firing pulses within the PRBS are staggered by a delay, which is set at least as long as the IGBT ignition or extinction delay, whichever is greater. D Flip Flops hold the switching state between the rising and falling edges of the PRBPT signal, with the new value used to set the state of each switch, hence the connection of switches 1 and 2 (positive voltage) to Q and 3 and 4 to !Q. To avoid disturbing integral controllers within the inverter, control states are frozen for the duration of probing. 

\section{Results}

Electro-magnetic transient (EMT) simulation, was carried out using the Simscape Specialized Power Systems toolbox. Simulations were run with a step size of 1 $\mu s$, consisting of a 0.05 second initialization period, a period for probing, and a total simulation time of 0.5 seconds.

An initial test was carried out in order to design the detector. The system was initialized, and each plant simultaneously probed the system. This was carried out for the intact system, along with all 12 possible switching states. Table \ref{nu_gap_table} shows the computed gap metric between the intact state and each possible open breaker, as seen the three solar plants.

 \begin{table}[h] 
    \centering
            \caption{Nu-Gap Distance to Intact System}
    \begin{tabular}{||c|c|c|c||}
    \hline
        Open Breaker & Plant 1 & Plant 2 & Plant 3  \\
           \hline
            1 & $\mathbf{0.981}$ & $\mathbf{0.989}$ & $\mathbf{0.977}$\\
             \hline
             2 & 0.068 & 0.039 & 0.116677\\
             \hline
             3 & 0.082 & $\mathbf{0.988}$ & $\mathbf{0.977}$\\
             \hline
             4 & 0.070 & 0.051 & 0.242\\
             \hline
             5 & 0.069 & 0.041 & 0.153\\
             \hline
             6 & 0.067 & 0.042 & 0.106\\
             \hline
             7 & 0.073 & 0.028 & $\mathbf{0.976}$\\
             \hline      
             8 & 0.071 & 0.044 & 0.491\\
             \hline             
             9 & 0.068 & 0.041 & 0.114\\
             \hline
             10 & 0.069 & 0.027 & 0.116\\
             \hline
             11 & 0.072 & 0.041 & $\mathbf{0.974}$\\
             \hline
             12 & 0.068 & 0.039 & 0.149\\
             \hline
    \end{tabular}
    \label{nu_gap_table}
    \vspace{-2mm}
\end{table}

The switching states that leave the plant islanded per Fig. \ref{Piaquadio_3} are bold. The islanded conditions produce a nu-gap metric near 1, while grid-connected conditions are much lower. This makes intuitive sense, as many controllers are expected to be unstable in islanded operation. As system conditions change, the high nu-gap of the islanded states also guarantees robust performance. A threshold-based detector is implemented by probing the system and computing the nu-gap against the stored realization for the intact network. If the nu-gap is larger than 0.9, the plant is islanded.

To evaluate the performance of both schemes, 300 independent Monte-Carlo simulations are carried out. Each load is assigned a random scaling factor between 0.5 and 1.5 in each simulation, and a random breaker is opened after 0.16 seconds. Table \ref{Monte_Table} shows the performance of the islanding detector.

 \begin{table}[h] 
  \vspace{-3mm}
    \centering
            \caption{Islanding Detector Performance}
    \begin{tabular}{||c|c|c|c|c||}
    \hline
        Plant & Connected Runs & Accuracy & Islanded Runs & Accuracy  \\
           \hline
            1 & 278 & 100 \% & 22 & 100 \%\\
             \hline
            2 & 252 & 100 \% & 48 & 100 \%\\
             \hline
            3 & 197 & 100 \% & 103 & 100 \%\\
        \hline
    \end{tabular}
    \label{Monte_Table}
    \vspace{-0.1cm}
\end{table}

An alternative approach is to store the realization from the nominal and all 12 switching states, compute all nu-gap metrics, and classify the system as the state with the lowest gap. This attempts to perform the topology detection, or determining which switch has opened. Table \ref{Monte_Topology} shows the performance of the topology detector.

 \begin{table}[h] 
    \centering
            \caption{Topology Detector Performance}
    \begin{tabular}{||c|c|c||}
    \hline
        Plant &  Topology Accuracy & Islanding Accuracy  \\
           \hline
            1 & 11.1 \% & 100 \%\\
             \hline
            2 & 12.1 \% & 100 \%\\
             \hline
            3 & 21.7 \% & 100 \%\\
        \hline
    \end{tabular}
    \label{Monte_Topology}
    \vspace{-1mm}
\end{table}

From Table \ref{Monte_Table}, all three solar farms correctly identified their islanded/grid connected condition in all replications. In Table \ref{Monte_Topology}, topology accuracy represents the percentage of replications in which each plant classified the correct breaker as open, while islanding accuracy represents the percentage of replications in which each plant would make the correct islanding decision (islanded/grid-connected). Plant 3 had slightly better topology performance, but the topology detection accuracy is low overall, as expected based on the outcome in Table \ref{nu_gap_table}. 

The results show perfect islanding detection accuracy for this simulated scenario, but leave out several practical considerations. Namely, sensor noise is not considered. Noise can lead to a poor signal to noise ratio in the probing output, worsening performance. Increasing the order of the probing signal can help mitigate noise impacts at the cost of increased probing time. Further, while only a single system is considered in this paper, distribution networks vary greatly in cabling, load composition, and topology. Consideration of noise and the exploration of varied systems is left as future work.

\section{Conclusions}

This work develops an active islanding detector using PCP. The method is rapid, with a longest probing duration of 223 ms, able to run on multiple DER simultaneously, and robust against load variation, as demonstrated in simulation. A physical implementation is provided via the logic of a switching inverter model. The methodology under performs as a topology detector, a problem which may be tackled via a multiple-model filter and making use of modern relay capability \cite{Wu_2018}.

The results so far only consider a single test system without noise, and the case of an exactly balanced islanded is left as future work. Further future work will include finding the precise non-detection zone for this method,  mathematically quantifying how much load or system model uncertainty can be tolerated before the detector fails.

\vspace{0mm}
\bibliography{bib}{}
\bibliographystyle{ieeetr}

\end{document}